\documentstyle{europhys}

\newif\ifboo \boofalse

\def\Review#1{\boofalse{\it #1},}
\def\Name#1{{\sc #1},}
\def\Vol#1{\ifboo Vol. {\bf #1}\else{\bf #1}\fi}
\def\Year#1{\ifboo #1\else(#1)\fi}
\def\Book#1{\bootrue{\it #1},}
\def\Page#1{\ifboo {\rm p. #1}\else{\rm #1}\fi}

\input{epsf.sty}
\begin{document}

\title{Chaotic dynamics in shear-thickening surfactant
       solutions}
\author{Ranjini Bandyopadhyay and A. K. Sood}
\institute{Department of Physics, Indian Institute of
           Science, Bangalore 560 012, India}

\maketitle

\begin{abstract}
We report the observation of dynamical behaviour in dilute,
aqueous  solutions of a surfactant CTAT (cetyl trimethylammonium
p-toluenesulphonate), below the overlap concentration c$^{\star}$.
At these concentrations, CTAT forms cylindrical micelles and shows
a pronounced shear thickening transition above a
concentration-dependent critical shear rate $\dot\gamma_{c}$. An
analysis of the time-series of the stress relaxations at
controlled shear rates in the shear-thickening regime shows the
existence of correlation dimensions greater than two and positive
Lyapunov exponents. This indicates the existence of deterministic
chaos in the dynamics of stress relaxation at these concentrations
and shear rates. The observed chaotic behaviour may be attributed
to the stick-slip between the shear - induced structure (SIS)
formed in the sheared surfactant solution and the coexisting
dilute phase. At still higher shear rates, when the SIS spans the
Couette, there is a transition to higher-dimensional dynamics
arising out of the breakage and recombination of the SIS.

\end{abstract}

\section{Introduction}
It is well known that surfactant molecules, under appropriate
conditions, can assemble reversibly in solution to form various
supramolecular structures. In very dilute solutions, at
concentrations above the critical micelle concentration, these
molecules self-assemble to form spherical micelles. Increasing the
surfactant concentration results in the formation of cylindrical
structures, which overlap and entangle to form clear viscoelastic
solutions of giant wormlike micelles. At still higher
concentrations, the surfactant solutions form mesophases
exhibiting nematic, hexagonal or lamellar ordering. It is possible
to pass from one phase to the other by simply changing the
surfactant and cosurfactant concentrations, the salinity or the
temperature of the system. The cylindrical structures that are
formed at very low concentrations $c$ ($c< c^{\star}$, $c^{\star}$
is the overlap concentration), get entangled on shearing at a
shear rate $\dot\gamma > \dot\gamma_{c}$ to exhibit a pronounced
increase in viscosity \cite{ch4gam,ch4liu,ch4hu}. At $\dot\gamma <
\dot\gamma_{c}$, the surfactant solutions exhibit Newtonian flow
behaviour. Recently, direct imaging studies of light scattered by
the shear-thickening CTAB/ NaSal and TTAA/ NaSal have shown the
existence of unstable {\it shear induced structures/ phases} (SIS/
SIP) \cite{ch4liu,ch4hu}, which fracture at very high shear rates,
giving rise to shear thinning \cite{ch4hu}. Direct imaging of the
SIS using freeze-fracture electron microscopy shows that these
structures span hundreds of microns and have sponge-like
textures\cite{ch4keller}.

In this paper, we present an experimental study on the nonlinear
dynamics of stress relaxation in dilute solutions of the
surfactant CTAT, at concentrations below the overlap concentration
c$^{\star}$. At these concentrations, CTAT forms cylindrical
micelles \cite{soltero1} and undergoes a shear-thickening
transition when sheared at $\dot\gamma > \dot\gamma_{c}$, which is
followed by shear-thinning at higher values of $\dot\gamma$
\cite{ch4gam}. The phase diagram and flow behaviour of aqueous
solutions of CTAT in the linear and nonlinear regimes have been
investigated extensively by Soltero {\it et al.} \cite{soltero1}.
Previous investigations of the flow behaviour of shear-thickening
solutions of CTAT \cite{ch4gam} using rheometry and neutron
scattering have shown the coexistence of a highly viscoelastic
shear-induced phase (SIP) with a viscous regime made up of short
aggregates. Oscillations in the apparent viscosity (and hence the
stress) of dilute CTAT at controlled shear rates in the shear
thickened regime have been observed by Gamez-Corrales {\it et al.}
\cite{ch4gam}. Observations of viscosity oscillations in
CTAB/NaSal \cite{ch4liu} and TTAA/NaSal \cite{ch4hu} in water
under controlled stress conditions have been explained in terms of
the growth and retraction of the SIS that form in the sheared
solution.

In a recent paper \cite{ran}, we have established the existence of
deterministic chaos \cite{ott} in the stress relaxation of
shear-thinning CTAT solutions at concentrations $c > c^{\star}$.
The oscillations in the normal and viscoelastic stresses were
explained in terms of the stick-slip between shear bands
\cite{ch4spe} that form in the plateau region of the flow curve.
Here, we extend our experiments to more dilute solutions of CTAT
($c$ = 11mM, zero-shear viscosity $\eta_{\circ} \sim$ 2mPas,
$c^{\star} \sim $ 15mM) and show that the occurence of coexisting
phases in a sheared surfactant solution is a prerequisite for the
existence of dynamical behaviour in its stress relaxation. In the
following sections, we will demonstrate the existence of
deterministic chaos in the stress relaxation of shear-thickened
CTAT, possibly due to the stick-slip between the SIS with the
coexisting dilute phases. To our knowledge, this is the first
observation of deterministic dynamics in the stress relaxation of
shear-thickening surfactant systems. The observed chaotic
behaviour is followed by an increase in complexity of the dynamics
of stress relaxation at higher shear rates due to the fracture of
the SIS, which is accompanied by the formation of large vortices
in the sheared solution \cite{ch4hu}.

\section{Sample Preparation and Experimental Apparatus}
CTAT, purchased from Sigma Chemicals, Bangalore, India was used to
prepare the samples by dissolving requisite amounts of CTAT in
doubly distilled and deionised water. The sample containers were
sealed and kept at 25$^{\circ}$C and shaken frequently to
accelerate homogenisation. By measuring the viscosities of CTAT
solutions prepared at different concentrations as a function of
the shear rate $\dot\gamma$, and extrapolating the values to
$\dot\gamma$=0, we have estimated the overlap concentration
$c^{\star}$ of CTAT in water to be $\sim$ 15mM. The rheological
measurements were made using a Rheolyst AR1000N stress-controlled
rheometer at T=25$^{\circ}$C, in a concentric cylinder geometry
with an conical end (outer cylinder diameter = 30mm, inner
cylinder diameter = 27.66mm, immersed height = 25.8mm).

\section{Experimental Results}

Figure 1 shows the flow curve of 11mM (0.5wt.\%) CTAT at
T=25$^{\circ}$C, measured by allowing 30 seconds between the
acquisition of successive data points. The flow curve shows all
the distinct regimes seen in TTAA/NaSal by Hu {\it et al.} in
stress-controlled experiments \cite{ch4hu}, {\it viz.}, a
Newtonian region of constant viscosity (Regime I) followed by a
shear thickening regime where the SIS nucleates heterogeneously
from the inner wall of the Couette cell (Regime II). In regime II,
the shear rate is seen to decrease with stress for a controlled
stress measurement, which supports the idea of coexistense between
the SIS and the dilute phase at these shear rates \cite{ch4gam}.
Next, we observe a shear-thickened regime where the stress
increases almost linearly with shear rate (Regime III). In this
regime, the SIS forms a percolating cluster that undergoes plug
flow under shear and nucleates homogeneously in the sheared
solution \cite{ch4hu}. Regime IV is a shear thinning regime
arising out of the  fracture of the SIS at very high shear rates.

Figure 2 shows the shear stresses measured on application of step
shear rates in the regimes II, III and IV. The shear stresses show
an initial increase followed by a chaotic time-dependence about a
mean value after a time $\tau_{p}$. We would like to point out at
this stage that the flow curve shown in Fig. 1 is {\it not} the
true steady state flow curve and the data in regimes II and III
are in all probability, transient. We come to this conclusion as
we find from Fig. 2 that steady state is not reached even after
shearing for 1500 seconds. We find that $\tau_{p} \sim \dot\gamma
^{-\alpha}$, where $\alpha$=1.8, in contrast to $\alpha$=1 for
dilute TTAA/NaSal solutions \cite{ch4hu}. The emphasis of this
work is on the analysis of the acquired time-series using standard
non-linear dynamical methods. We prove the existence of
deterministic dynamics in the stress relaxation of sheared CTAT
for shear rates lying in regimes II and III, where we have derived
finite values of the correlation dimensions and positive values of
the Lyapunov exponents. The dynamics in regime IV shows a
transition to higher degrees of complexity due to the stronger
spatial dependence of the flow fields and a consequent coupling of
the different wavevector modes, arising out of the percolation and
fracture of the SIS.
\section{Data Analysis}
We have employed the standard tools of nonlinear dynamics
\cite{ott} to calculate certain metric and dynamical invariants
\cite{abarbanel} characterising the time-series measured in our
experiments and displayed in Fig. 2. This is done to distinguish
whether our signals arise out of stochastic or deterministic
processes. As opposed to stochastic noise, chaotic dynamics may be
predicted over short time scales and are extremely sensitive to
initial conditions. The first invariant we have calculated is the
{\it correlation dimension} $\nu$ which gives us an idea of the
geometry of the attractor in phase space on which the stress
trajectories lie in the asymptotic limit. Next, we have calculated
the largest {\it Lyapunov exponent} $\lambda$ which provides a
measure of the divergence of neighbouring trajectories. Presence
of chaotic dynamics in a system requires $\nu >$ 2 and $\lambda >$
0. The analysis is done by reconstructing the phase space by
embedding the experimental time-series of shear stress $\sigma$ in
$m$ dimensions using time delay vectors $L$ \cite{abarbanel}, so
that the invariants associated with the dynamics may be calculated
unambiguously. Let $\sigma_j = \sigma (j\Delta t)$ denote the
time-series shown in Fig. 2, consisting of stresses measured at
regular time intervals $\Delta$t (=1s), with $j$=1 to N (N=1500),
at various shear rates maintained constant during the duration of
the experiment. A reasonable value of $L$ ( =$L_{\circ}$) has been
estimated for each time-series by plotting the stress
autocorrelation function $C(\tau)$ versus $\tau$. The value of
$\tau$ at which $C(\tau)$ falls to $\frac{1}{e}$ of its maximum
value of 1 (at $\tau$ = 0) is used as an estimate of the optimum
delay time $L_{\circ}$. An optimal value of the minimal embedding
dimension $m_{\circ}$ has been estimated using the method of false
nearest neighbours proposed by Kennel {\it et al.}
\cite{ch4heg,ch4tis}. The values of $L_{\circ}$ thus obtained are
used to construct the sequence of vectors \{{$\vec
X_i:i$=1,2,....N- ($m$-1)$L_{\circ}$}\} which describe
trajectories in $m$-dimensional phase space arising out of the
dynamics $F: \vec X_i \rightarrow \vec X_{i+1}$, and which are
used for the computation of the invariants characterising the
dynamics such as the correlation dimension $\nu$ and the Lyapunov
exponent $\lambda$ \cite{abarbanel}. The calculations of the
correlation dimension $\nu$ uses the  algorithm of Grassberger and
Procaccia \cite{grassberger} which involves the calculation of the
correlation integral $C(R)$. The correlation integral C(R) is
defined in an $m$-dimensional phase space as $C(R)={lim\atop
N\rightarrow \infty }{1\over N^{2}}\sum _{i,j=1}^{N}H({R-|\vec
X}_{i}-\vec X_{j}|)$, where H(x) is the Heaviside function and
$|\vec X_i-\vec X_j|$ is the distance between the pair of points
$(i,j)$ in the $m$-dimensional embedding space. The sum in the
expression for $C(R)$ gives the number of point pairs separated by
a distance less than R. For small $R$s, $C(R)$ is known to scale
as $C(R)\sim R^{\nu }$, where $\nu$ is the correlation dimension
\cite{grassberger}. A plateau in the plot of $\nu$ versus
$\log(R)$ gives the correct $\nu$ for a chosen embedding dimension
m. If the attractor is unfolded by choosing a large enough $m$,
then the correlation dimension $\nu$ becomes independent of the
value of the embedding dimension $m$. The value of $m = m_{\circ}$
for which this independence sets in is the correct embedding
dimension, and the corresponding $\nu$ is the correlation
dimension. If an experimental signal satisfies $\nu < m$, then the
signal is due to deterministic chaos rather than random noise
\cite{grassberger}. Figure 3 (a) shows a plot of $\nu$ versus
$\log(R)$ for $m$=1-6 (increasing in the direction shown by the
arrow) at $\dot\gamma$=60s$^{-1}$, which corresponds to regime II
of the flow curve. The value of $\nu$, obtained from the plateau
in the $\nu$ versus $\log(R)$ plot, is found to saturate to a
value of 2.6, which gives $m_{\circ} \sim$ 3 or 4. The inset shows
the values of the correlation dimensions $\nu$ versus the
embedding dimensions $m$ calculated from the experimental data
(open circles, marked as E) and five sets of surrogate data
(SI-SV) generated by randomising the phases of the original data
\cite{ch4heg}. The values of $\nu$ calculated from the surrogate
data do not converge even at $m$=6 and lie well above the values
calculated from the experimental time-series. This implies the
presence of deterministic dynamics in the stress relaxation of
shear-thickening CTAT at $\dot\gamma$=60s$^{-1}$.

Figure 3(b) shows the plot of $\nu$ versus $\log(R)$ for the data
set acquired at $\dot\gamma$=400s$^{-1}$ which saturates to an
approximate plateau value of 4.5 for $m\ge$6 over a range of
$\log(R)$. The values of $\nu$ calculated from the experimental
and surrogate time-series are not very different in this case.
These large values of $\nu$ and $m_{\circ}$ calculated for
$\dot\gamma$ = 400s$^{-1}$ may be interpreted as a manifestation
of the increased complexity of the dynamics of the stress
relaxation in regime IV arising out of the fracture of the SIS
which is accompanied by the formation of vortices in the sheared
solution.   Figure 4 shows a plot of $\nu$ versus $\dot\gamma$ for
the regimes II, III and IV. Figure 5 shows the calculation of the
largest Lyapunov exponent for  $\dot\gamma$=40, 50 and 90s$^{-1}$,
following the algorithm of Gao and Zheng \cite{gao}. This is done
by defining $d_{ij}(k)=||{\vec X_{i+k}-\vec X_{j+k}}||$, the
Euclidean distance between two vectors constructed using the
embedding theorem by $k$ iterations of the dynamics F, and
plotting $\Lambda$=$<\ln[d_{ij}(k)/d_{ij}(0)]>$ as a function of
$k$. The Lyapunov exponents, calculated by using the relation
$\lambda = S/\Delta t \ln(2)$, where S is the slope of the plot,
are found to be positive at all shear rates in regimes II and III,
which confirms the existence of deterministic chaos at these shear
rates. Due to the limited  extent of our time-series, we have been
unable to obtain reliable values of $\lambda$ for shear rates
lying in regime IV. Preliminary analysis indicates an increase in
the values of $\lambda$ in regime IV from those obtained in regime
III, confirming the existence of higher dimensional chaos in this
regime. The inset of Fig. 5 shows the values of $\lambda$ at
different shear rates in  regimes II and III.

\section{Conclusions}
In summary, we have proved the existence of deterministic chaos in
the stress relaxation of dilute CTAT solutions in the
shear-thickening regimes (Regimes II and III). The chaotic
dynamics can arise out of the stick-slip between the SIS and the
coexisting dilute phase in the sheared solution. In regime IV,
where the spatial dependence of the flow fields becomes
significant due to the percolation and fracture of the SIS, we
have demonstrated an increase in the complexity of the dynamics,
which is reflected by an increase in the optimal embedding
dimension $m_{\circ}$, the correlation dimension $\nu$ and the
Lyapunov exponent $\lambda$. Stress relaxation in dilute CTAT,
therefore, provides yet another example of deterministic chaotic
dynamics in a driven system arising out of the coupling between
the lateral and normal degrees of freedom \cite{ran,ch4roz}. We
hope that our work will stimulate theoretical understanding of the
nonlinear flow of dilute surfactant solutions in terms of
nonlinear coupled equations having additional predictive
capabilities.

\section{Acknowledgements}
We thank Sriram Ramaswamy, P. R. Nott and V. Kumaran for the use
of the rheometer.




\newpage
FIGURE CAPTIONS \\

Fig. 1. - (a) shows a plot of the shear stress $\sigma$ while (b)
shows a plot of the viscosity $\eta$ versus shear rate
$\dot\gamma$. Depending on the behaviour of $\eta$ at different
shear rates, the flow curve may be divided into four regimes :
regime I where the surfactant solution shows Newtonian flow
behaviour, regime II characterised by shear-thickening as a result
of the nucleation of SIS, regime III where the SIS spans the
Couette and regime IV which shows shear-thinning as a result of
the fracture of the SIS. The filled circles represent the
shear-rate controlled measurements while the open squares denote
stress control. \\

Fig. 2. - Plots of the relaxation of the shear stress $\sigma$  at
shear rates $\dot\gamma$= (a) 40, (b) 50, (c) 60 s$^{-1}$(regime
II), (d) 90, (e) 200 s$^{-1}$(regime III) and (f) 400s$^{-1}$
(regime IV). \\

Fig. 3. - The calculations of the correlation dimensions from the
time-series of stresses at $\dot\gamma$= (a) 60$^{-1}$ and (b)
400s$^{-1}$. The insets show the calculations of $\nu$ for the
experimental data set (denoted by E) and five sets of surrogate
data ( SI-SV) generated according to the algorithm of Theiler {\it
et al.} \cite{ch4tis}. $\nu$ saturates to 2.6 for the experimental
data at $\dot\gamma$ = 60$^{-1}$. At $\dot\gamma$ = 400$^{-1}$,
the $\nu$'s calculated from the experimental data and the
surrogate data sets are not very different, which indicates a
transition to higher-dimensional dynamics.\\

Fig. 4. - Plot of $\nu$ versus $\dot\gamma$, which shows a
monotonic increase of $\nu$ versus $\dot\gamma$.\\

Fig. 5. - Calculations of $\Lambda$ versus $k$ for
$\dot\gamma$=40, 50 and 90s$^{-1}$. The Lyapunov exponents
$\lambda$ which may be estimated from the slopes of the plots, are
found to increase with $\dot\gamma$, as shown in the inset.\\

\newpage
\begin{figure}
\centerline{\epsfxsize = 15cm \epsfysize = 18cm
\epsfbox{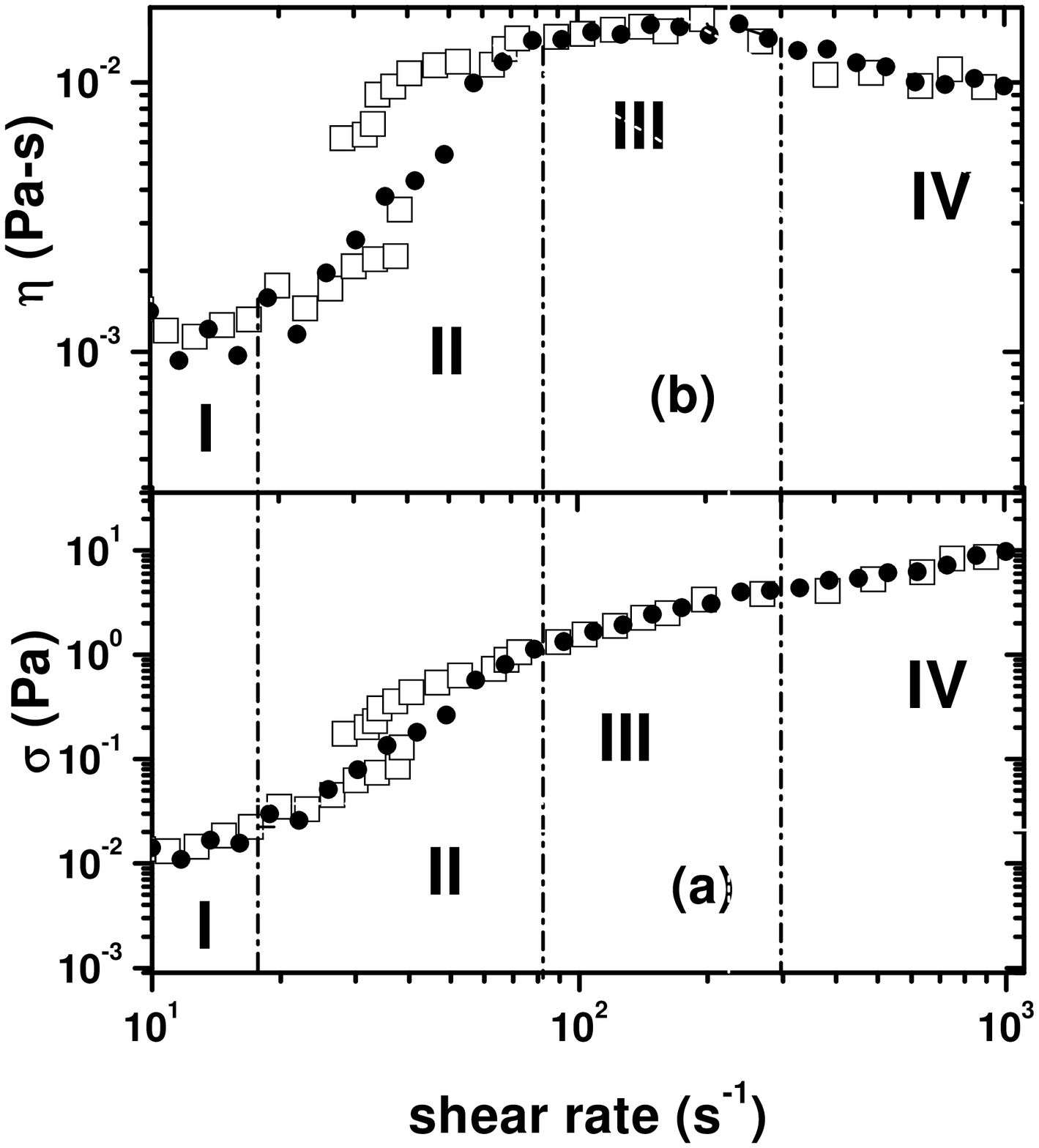}}
\end{figure}
\centerline{\bf Figure 1}

\newpage
\begin{figure}
\centerline{\epsfxsize = 15cm \epsfysize = 18cm
\epsfbox{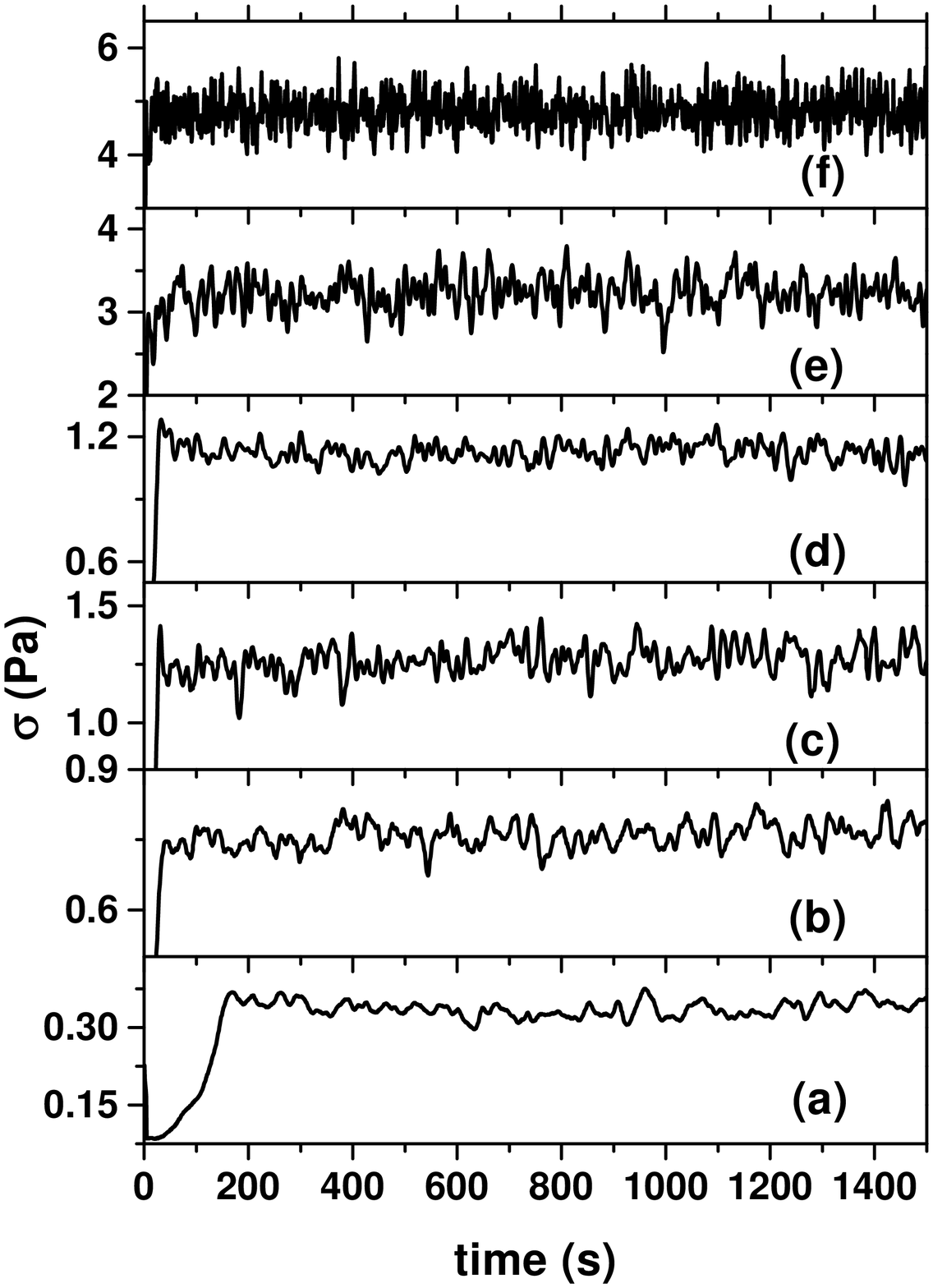}}
\end{figure}
\centerline{\bf Figure 2}

\newpage
\begin{figure}
\centerline{\epsfxsize = 16cm \epsfysize = 13cm
\epsfbox{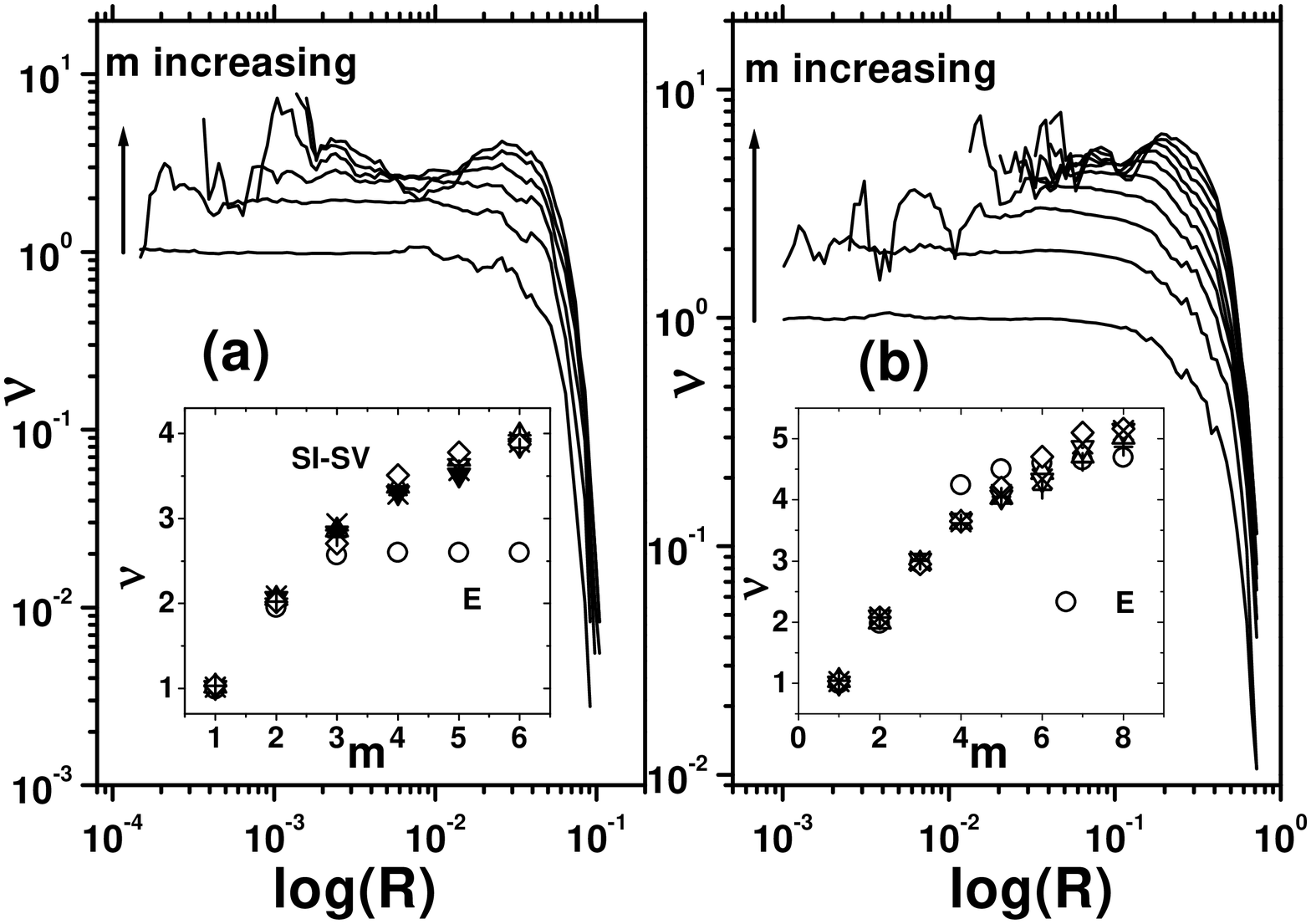}}
\end{figure}
\centerline{\bf Figure 3}

\newpage
\begin{figure}
\centerline{\epsfxsize = 12cm \epsfysize = 15cm
\epsfbox{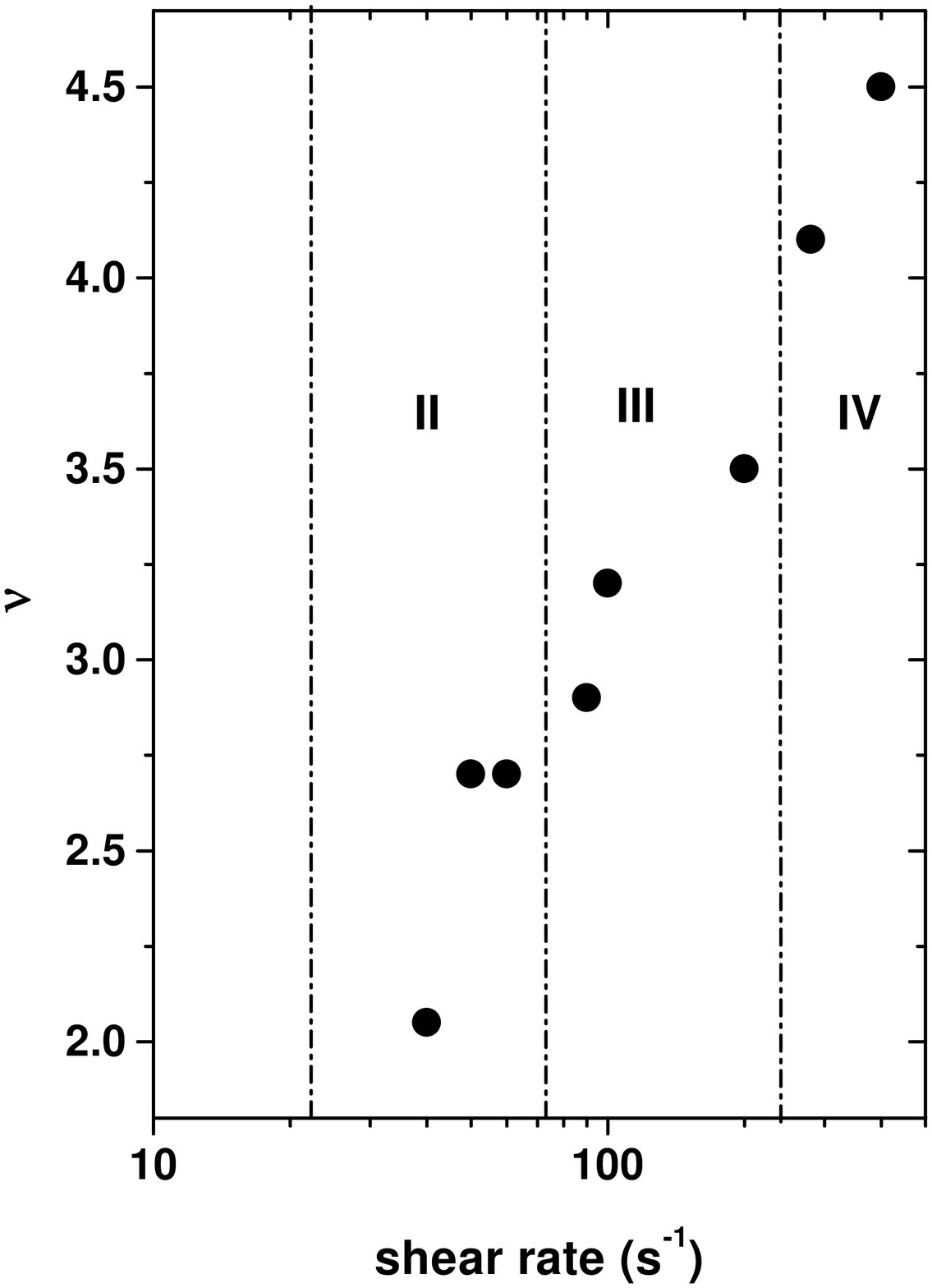}}
\end{figure}
\centerline{\bf Figure 4}

\newpage
\begin{figure}
\centerline{\epsfxsize = 13cm \epsfysize =
15cm\epsfbox{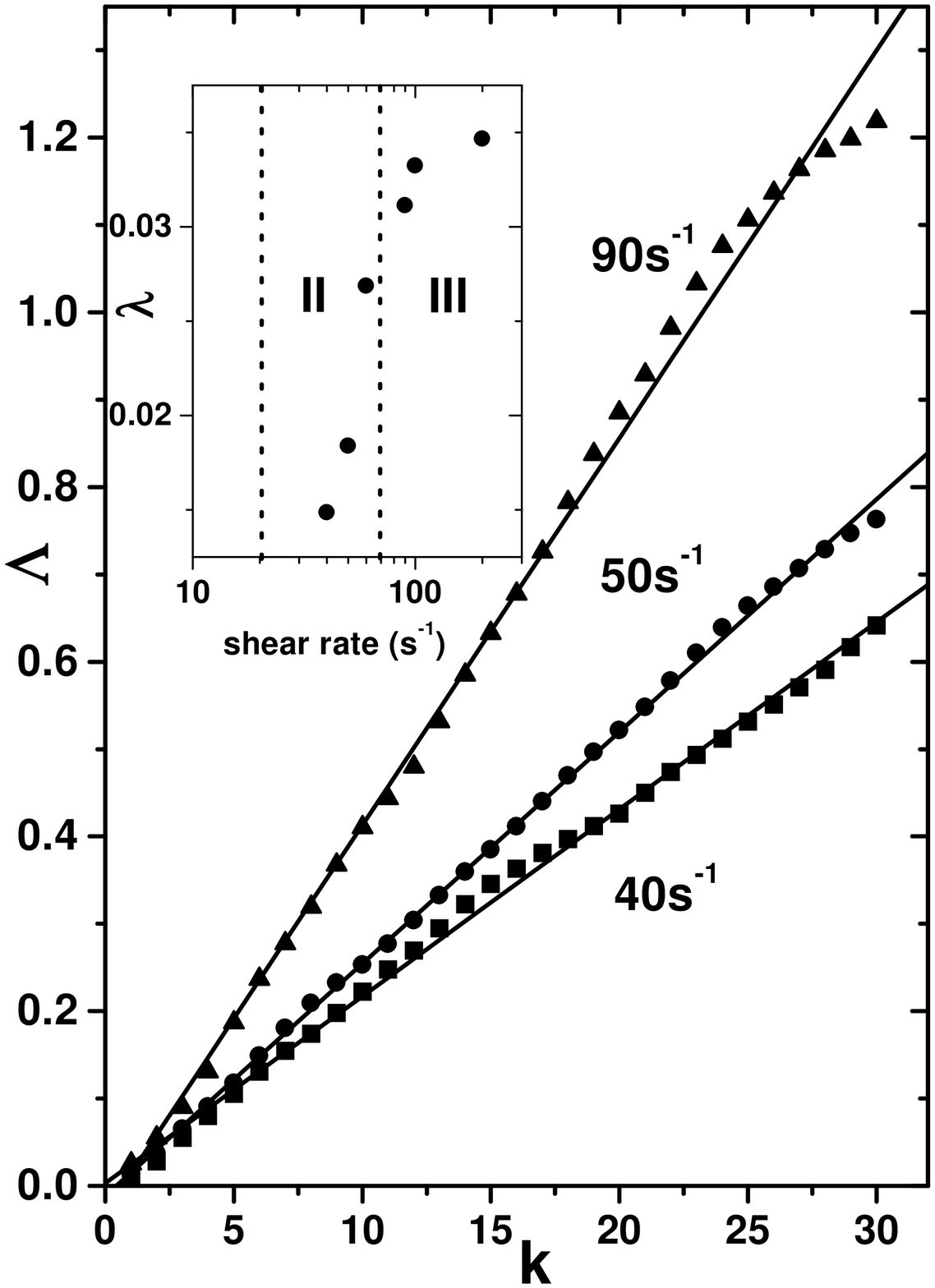}}
\end{figure}
\centerline{\bf Figure 5}

\end{document}

\end{document}